\documentclass{article}

    \usepackage[preprint]{neurips_data_2024}
\usepackage{graphicx}
\usepackage[utf8]{inputenc} 
\usepackage[T1]{fontenc}    
\usepackage{hyperref}       
\usepackage{url}            
\usepackage{booktabs}       
\usepackage{amsfonts}       
\usepackage{amssymb}
\usepackage{nicefrac}       
\usepackage{microtype}      
\usepackage{xcolor}         
\usepackage{lipsum}
\usepackage{xspace}
\usepackage{longtable}
\usepackage{lscape}
\usepackage{subcaption}
\usepackage{amsmath}
\usepackage{multirow}
\usepackage{pifont}
\usepackage[table]{xcolor} 
\usepackage[hang,flushmargin]{footmisc}

\let\oldding\ding
\renewcommand{\ding}[2][1]{\scalebox{#1}{\oldding{#2}}}

\usepackage{siunitx} 
\usepackage{caption} 

\newcommand{\name}{\textsc{DroneAudioNet}\xspace}

\newcommand{\second}[1]{\underline{#1}}
\newcommand{\sigft}{^{\dagger}}

\newcommand{\cmark}{\ding[1.2]{51}}%
\newcommand{\xmark}{{\ding[1.2]{55}}}%

\title{\name: Noise Suppression for Drone Audition-based Search and Rescue}
\author{
  \normalfont Chitralekha Gupta\textsuperscript{*} \and Soundarya Ramesh\textsuperscript{*} \and Yifei Luo \and Suranga Nanayakkara\\
  \textsuperscript{*}Equal contribution \\
  School of Computing, National University of Singapore \\
  \texttt{\{chitralekha, soundarya, yifei, suranga\}@ahlab.org}
}

\begin{document}

\maketitle

\begin{abstract}
Microphones mounted on UAVs enable aerial acoustic scene analysis applications such as search-and-rescue, wildlife monitoring, and industrial inspection. 
However, drone rotor noise often dominates the mixture signal at SNRs well below -10 dB, making source recovery extremely challenging. Existing enhancement and source separation methods are typically designed for near-balanced mixtures and degrade substantially in drone audition settings. In this work, we propose \name, a drone noise suppression method that reframes a source separation model as a drone noise estimator. To better model drone-dominant mixtures, we introduce a learnable mask-scaling mechanism that allows mask magnitudes beyond unity, together with an additive residual correction term for improved drone estimation and source recovery. We train and evaluate our model on a publicly available drone audition dataset and test generalizability on an out-of-domain dataset with unseen drone hardware and flight modes. 
Results show that \name consistently improves downstream sound classification performance, with the largest gains observed for human vocal sounds. 
Our findings demonstrate the importance of drone-specific modeling for robust aerial acoustic perception and highlight the potential of source separation methods for real-world drone-assisted search-and-rescue.

\end{abstract}

\vspace{-0.2cm}
\section{Introduction}
\label{sec:introduction}
\vspace{-0.2cm}
Microphones mounted on UAVs open up a range of aerial acoustic scene analysis applications, including search-and-rescue, wildlife monitoring, and industrial inspection, by providing an elevated acoustic vantage point. The main challenge, however, is the drone itself: rotor blades generate intense, time-varying noise comprising harmonics of the blade-passing frequency and broadband turbulence, often overwhelming ambient sounds of interest by 10--20 dB~\cite{deleforge2020drone,martinez2020review}. Existing enhancement and separation approaches, whether speech enhancement pipelines~\cite{lu2023mp} or universal source separation models such as USS~\cite{kong2023universal} and AudioSep~\cite{liu2024separate}, are primarily developed for near-balanced mixtures (typically SNR $>-5$ dB) and degrade substantially in these extreme low-SNR conditions.

A key property of this problem, however, works in the practitioner's favour: unlike generic noise suppression, the interference type is known. Rotor noise has a well-defined physical origin, structured harmonic content, and characteristic phase evolution~\cite{deleforge2020drone}. Rather than estimating an unknown target source directly, we can instead explicitly model the drone noise and subtract it from the mixture, leaving the residual ambient signal intact regardless of its semantic category. This formulation is especially attractive in drone audition, where the target sounds may be open-domain and highly heterogeneous, ranging from speech and screams to environmental and mechanical events.


While recent source separation models such as AudioSep~\cite{liu2024separate} demonstrate strong performance across diverse audio domains, directly applying them to drone noise suppression remains challenging. Unlike typical source separation settings where sources have comparable energy, drone audition often involves extremely low-SNR mixtures in which drone noise dominates weak target sounds. 
In such conditions, conventional mask-based separation methods can systematically underestimate the drone component due to their bounded mask formulation~\cite{kong2023universal, chen2022zero, liu2024separate}. To address this, we propose \name, which reframes AudioSep as a drone noise estimator, 
and introduces architectural modifications tailored to the acoustic characteristics of drone noise. Specifically, \name incorporates a learnable mask-scaling mechanism that allows mask magnitudes beyond unity, along with an additive residual correction term for improved drone estimation and source recovery. 


\name achieves significant gains through its relaxed mask parameterization and additive residual correction, with the largest improvements (10.6\% relative increase in source classification  over the best baseline) observed for human vocal sounds under severe drone interference (SNR between -20 to -10 dB).  
Evaluations on the out-of-domain dataset demonstrate that \name generalizes effectively to unseen drones and flight modes, particularly for vocal 
sounds such as speech and cries. 

We make three contributions. First, we propose \name, by adapting  state-of-the-art source separation method, AudioSep. Second, we provide a comprehensive benchmark for universal source separation methods on a publicly available drone dataset\footnote{Links to Github codebase and webpage consisting of sample audio files are available in the Appendix~\ref{app:webpage}.}. Third, we report improvements in signal fidelity and downstream source classification performance on in-domain and out-of-domain datasets. Our work is a step towards real-world drone-assisted search and rescue.

\vspace{-0.2cm}
\section{Related Work}
\label{sec:bg-and-related-work}
\vspace{-0.2cm}
\subsection{Drone Noise Suppression}
Prior work on drone audition and UAV acoustics has explored a range of hardware-assisted and classical signal-processing approaches \cite{martinez2020review}. 
Representative systems include spherical or phased microphone arrays for source localization \cite{manamperi2022drone,manamperi2024drone}, direction-of-arrival (DOA) estimation with beamforming \cite{go2021acoustic,manamperi2022drone}, 
Wiener/post-filtering and adaptive filtering pipelines \cite{serrenho2019gunshot,manamperi2024drone,minea2023urban} for specific tasks such as search and rescue, gunshot surveillance, traffic-noise mapping, and bioacoustic monitoring. However, evaluations are typically conducted under task-specific setups and datasets, making reproducibility and comparison difficult. For example, Drone Audition~\cite{manamperi2022drone} evaluates localization down to approximately $-30$~dB SdNR using a hovering drone in a semi-anechoic chamber with a 30-channel microphone array, while the gunshot detection work of AIRA-UAS~\cite{ruiz2018aira} reports experiments at 0, 2, 5, and 10~dB SNR. 
Moreover, while a few works released datasets or relied on public toolkits, such as DREGON \cite{strauss2018dregon}, AIRA-UAS \cite{ruiz2018aira}, and AVQ \cite{wang2019audio}, 
most did not release public code or pretrained models, or the data is not large and diverse enough for training a dedicated drone noise suppression pipeline. This limits direct comparison and makes it difficult to reuse earlier systems across new drone platforms and recording conditions, making it difficult to establish standardized baselines for drone-noise suppression. 


\subsection{Low-SNR Audio Enhancement and Universal Source Separation}
\label{sec:related-work-enhancement}

Audio enhancement and source separation aim to recover signals of interest from noisy mixtures, typically by estimating time--frequency (TF) masks (magnitude and phase) with respect to the \emph{target source}~\cite{luo2019convtasnet, subakan2021attention, lu2023mp}. This target-centric formulation assumes that the signal of interest is known \emph{a priori} and sufficiently represented during training. In practice, a large body of work focuses on speech enhancement, where models are optimized for human speech characteristics (e.g., VoiceBank+DEMAND). 
However, these assumptions do not hold in heterogeneous acoustic environments. Our setting involves arbitrary, open-domain sounds (e.g., human vocalizations such as screams or crying, mechanical events, and environmental sounds), where the target distribution is not predefined. Consequently, speech-specific models do not generalize well, motivating the use of class-agnostic, universal audio separation approaches.

To address this, we build upon \emph{universal audio source separation} methods that are trained on large-scale, diverse audio corpora and generalize across sound categories. We consider three representative baselines. \textbf{Zero-shot separation}~\cite{chen2022zero} formulates separation as a query-conditioned task, enabling extraction of unseen sound classes using weak supervision. \textbf{Universal Source Separation (USS)}~\cite{kong2023universal} extends this paradigm to all AudioSet classes via a FiLM-conditioned ResUNet, providing strong coverage over diverse sound events. \textbf{AudioSep}~\cite{liu2024separate} further incorporates language conditioning through CLAP embeddings, enabling flexible text-guided separation and improved generalization. These models typically estimate the mask of the sound of interest indicated by the query. These models are well-suited as baselines in our setting due to their class-agnostic design, large-scale pretraining on a diverse set of sounds from AudioSet \cite{gemmeke2017audioset}, and demonstrated zero-shot capability across heterogeneous audio domains.

Despite these advances, existing audio enhancement and source separation models typically assume balanced mixtures of source and noise (SNR $\approx$ 0 dB). In contrast, drone audition operates in extreme low SNRs, altering the behavior of ideal masks and limiting the effectiveness of standard formulations.

\subsection{Explicit Noise Modeling and Subtraction Paradigms}
\label{sec:related-work-noise-modeling}

An alternative to target-signal estimation is to explicitly model and subtract the interference signal. This paradigm has been explored in speech enhancement, where neural networks predict noise components that are removed from the mixture in the time or TF domain~\cite{odelowo2017noise,zheng2021interactive}. Such approaches exploit the relatively stable structure of noise compared to target signals. These noise-prediction methods are developed for speech enhancement of signals with SNR >-5dB. On the other hand, mask-based advances, such as complex ideal ratio mask related modifications are largely studied for signals where the source is dominant \cite{kong2021decoupling}. Building upon these studies, we formulate our problem as known-noise estimation under extreme low-SNR conditions, combining explicit noise modeling with an unbounded complex mask representation tailored to noise-dominant mixtures.

\vspace{-0.2cm}


\section{Methods}
\label{sec:dronenet}
\name builds upon on the state-of-the-art source separation method, 
\textit{AudioSep}~\cite{liu2024separate}, by reframing it as a drone noise estimator (Section~\ref{sec:methods-reframing}), and subsequently updating its architecture to improve its suitability for modeling drone noise (Section~\ref{sec:methods-improved-mask-estimation}).

\begin{figure}
    \centering
    \includegraphics[width=1.0\linewidth]{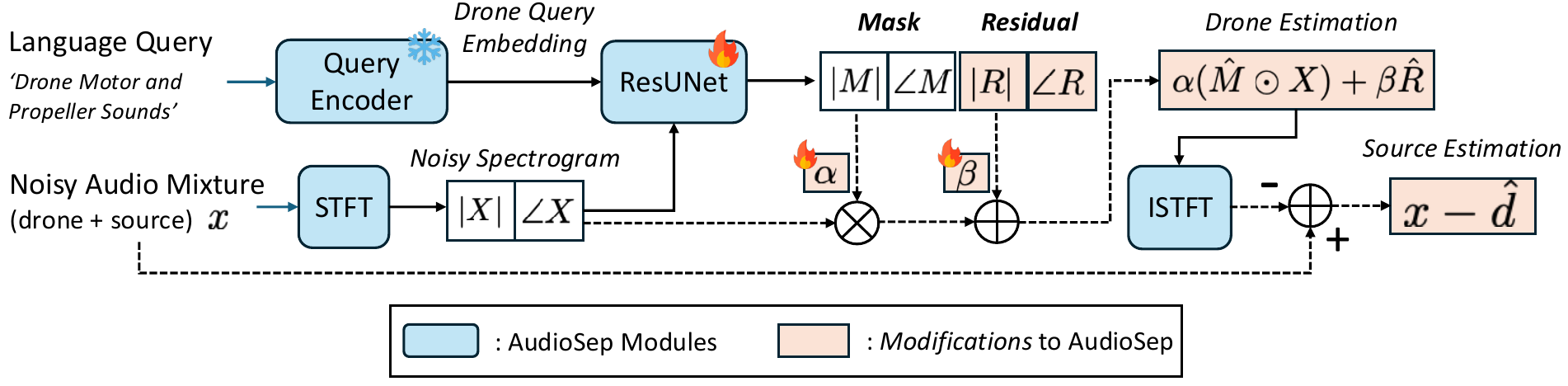}
    \caption{Overview of the \name architecture, adapted from AudioSep~\cite{liu2024separate}. To better estimate drone noise, we introduce a learnable mask-scaling parameter $\alpha$ for magnitudes beyond unity and an additive residual correction term $R$. The recovered source is obtained by subtracting the estimated drone signal from the mixture.}
    \label{fig:framework}
\end{figure}

\subsection{Reframing Source Separation Model as a Drone Noise Estimator}
\label{sec:methods-reframing}
We build our solution based on AudioSep~\cite{liu2024separate}, a natural language-query based universal source separation method, that performs the best in our benchmarking experiments (results presented in Section~\ref{sec:evaluation}). We cannot directly apply AudioSep to our problem as it requires a text query corresponding to the \textit{target source of interest}, which in our case, 
is \textit{unknown}. However, given that the noise source is \textit{known}, we use AudioSep with a fixed drone-based query, \textit{``Drone motor and propellor sounds''}, to estimate the drone noise, $\hat{d}$, from the noisy mixture, $\boldsymbol{x}$. Subsequently, we estimate the unknown source sound, $\hat{s}$, as the subtraction, $\hat{s} = \boldsymbol{x} - \hat{d}$, in the time-domain~\cite{odelowo2017noise, zheng2021interactive}. 

\begin{figure}
    \centering
    \includegraphics[width=0.7\linewidth]{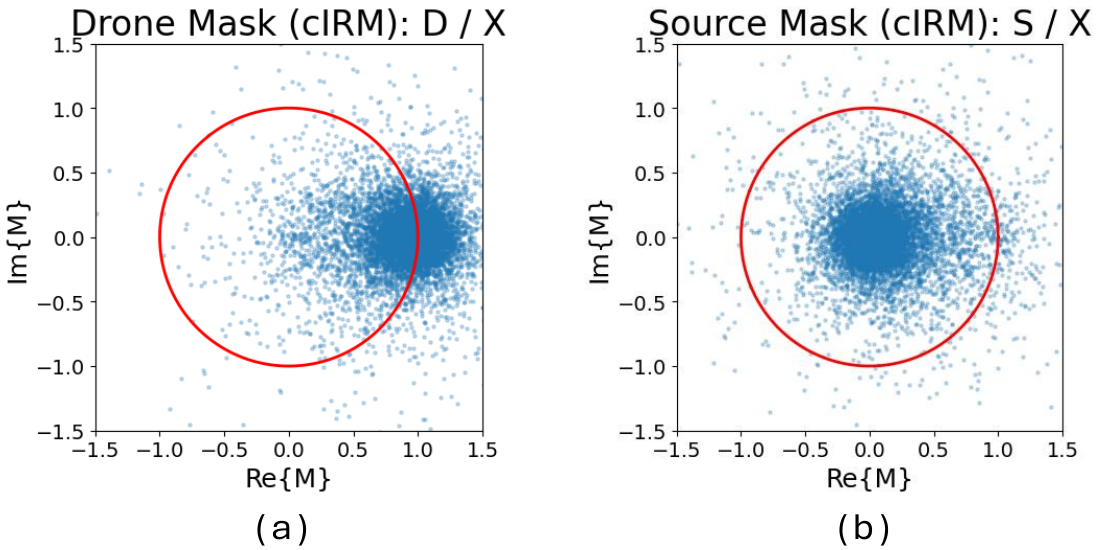}
    \caption{Complex ideal ratio mask (cIRM) distributions for (a) the source signal ($S/X$) and (b) the drone signal ($D/X$), with the unit circle shown in red. While only 0.25\% of source-mask values lie outside the unit circle, 49.3\% of drone-mask values exceed unity, motivating the need for unbounded mask magnitudes in drone noise estimation.} 
    \label{fig:cirm_comparison}
\end{figure}

\subsection{Improved Mask Estimation with Additive Residue}
\label{sec:methods-improved-mask-estimation}

We now describe how AudioSep estimates the queried drone component and motivate the architectural modifications introduced in \name. AudioSep uses a frequency-domain ResUNet separator~\cite{liu2024separate}, which takes the mixture spectrogram, $X$, and the query embedding as input, and predicts a magnitude mask $|M|$ and phase residual $\angle M$. The estimated drone spectrogram from AudioSep, $\hat{D}_{audiosep}$ is $\hat{D}_{audiosep} = \hat{M} \odot X = |M|\odot |X|e^{j(\angle X + \angle M)}$, 
where $\hat{M}$ is the estimated mask, and $|M|$ determines how much mixture energy is retained in each time-frequency bin, while $\angle M$ rotates the mixture phase, $\angle X$.

A key limitation of directly adopting this formulation for drone noise estimation is the \textit{bounded} magnitude-mask parameterization. In AudioSep, the magnitude mask is obtained through a \texttt{sigmoid} nonlinearity, constraining $|M|\in(0,1)$. This constraint is often useful in source-separation models because it regularizes the solution and stabilizes training. However, it implicitly assumes that the target source can be recovered by attenuating the mixture magnitude in each time-frequency bin, which can be restrictive for drone noise suppression. First, our target component is the drone noise rather than the weaker signal of interest; in many recordings, the drone dominates the mixture by over 10~dB. 
Second, when source signal, $S$, and drone signal, $D$, interfere destructively, the mixture magnitude $|X|=|S+D|$ can be smaller than the magnitude of either component, causing the ideal ratio $D/X$ or $S/X$ to exceed the unit circle~\cite{williamson2015complex,kong2021decoupling}. We show this in Figure~\ref{fig:cirm_comparison}, where the complex ideal ratio mask (cIRM) for drone noise, $D/X$, contains substantial mass outside the unit circle\footnote{For this illustration, we compute the ideal masks by combining a drone-noise recording with a human non-vocal source recording from DroneAudioSet~\cite{gupta2025droneaudioset} and taking the corresponding spectrogram ratios.}. 

To address this, \name modifies AudioSep in two complementary ways (Figure~\ref{fig:framework}). 

\noindent\textit{Relaxing the bounded mask through a learnable scale.}
We include a learnable scalar, $\alpha$ (initialized to 1), that is multiplied with the magnitude mask, that can scale the mask to values beyond 1, especially in drone-dominant mixtures and destructive-interference regions. We learn the scalar, $\alpha$,  
while keeping the magnitude mask, $|M|$, bounded through a \texttt{sigmoid} activation to maintain training stability. The updated estimate drone signal, $\hat{D_M}$, is $\hat{D}_{M} = \alpha (\hat{M} \odot X)$.  

\noindent\textit{Adding a residual spectrogram correction.}
We incorporate an additive residual term -- with magnitude, $\beta |R|$, and phase, $\angle R$, to enable direct drone signal prediction, similar to prior work~\cite{kong2021decoupling}. Such an additive formulation provides an additional degree of freedom, and serves as a corrective term that compensates for systematic underestimation in the mask. Similar to the magnitude mask, here, we learn $|R|$ with the \texttt{sigmoid} constraint, while scalar, $\beta$ 
(initialized to 1), enables learning of residual magnitudes exceeding unity. Overall, the estimated drone signal, $\hat{D}$, is $\hat{D} = \hat{D}_{M} + \beta \hat{R} = \alpha (\hat{M} \odot X) + \beta |R| e^{j\angle R}$, where $\hat{R}$ is the estimated drone residual. 
Together, the scaled mask and additive residual enable \name to adapt to acoustic characteristics of drone noise.

\vspace{-0.25cm}
\section{Evaluation}
\label{sec:evaluation}
\vspace{-0.2cm}
\subsection{Datasets}
\label{sec:eval-overview-datasets}

We use two public drone-audio datasets in this work. We use DroneAudioSet~\cite{gupta2025droneaudioset}, the largest publicly available drone audition dataset, for model fine-tuning and in-domain evaluation, and DREGON~\cite{strauss2018dregon} for out-of-domain (OOD) evaluation under unseen drone hardware and flight conditions.

\paragraph{DroneAudioSet.}
We use DroneAudioSet~\cite{gupta2025droneaudioset}, a publicly available drone audition dataset containing drone-only, source-only, and drone-with-source recordings across diverse recording conditions, including different drone platforms, microphone configurations, throttle levels, and acoustic environments, while the drone is hovering. Since our focus is on practically relevant low-SNR conditions, we retain configurations with input SNR $\geq -30$~dB, resulting in 101 recording configurations spanning (i) Human Vocal (HV) sounds which includes male and female speech/screams as well as infant crying sounds, (ii) Human Non-Vocal (HNV) sounds such as footsteps and clapping, as well as (iii) Non-Human (NH) sounds such as fire alarms and water trickling sounds. We split the retained data into train, validation, and test sets, resulting in 74.4, 24.7, and 25.3 hours of audio, respectively. We ensure that the source sounds across the three sets are mutually exclusive. As DroneAudioSet records each configuration simultaneously across multiple microphones, we treat each channel independently, resulting in a single-channel setup with increased spatial diversity. For training and validation, we synthetically construct mixtures by combining aligned drone-only and source-only recordings from the same configuration, whereas evaluation is performed directly on the naturally recorded drone-with-source mixtures to reflect realistic recording conditions.

\paragraph{Out-of-Domain Test Set.}
\label{sec:oodtestset}
To evaluate out-of-domain (OOD) generalization, we construct an OOD test set using drone-only recordings from DREGON~\cite{strauss2018dregon}, which differs substantially from DroneAudioSet in terms of recording setup and flight dynamics. DREGON includes five drone flight modes: hovering, free-flight, up-and-down, rectangle, and spinning. We combine these recordings with source-only recordings from DroneAudioSet to construct mixtures spanning input SNRs from $-13.7$~dB to $-19.5$~dB. Each microphone channel is treated independently as a mono recording, resulting in a 3.25-hour OOD test set covering the same sound categories as DroneAudioSet.
\subsection{Evaluation Metrics}
\label{sec:eval-overview-metrics}

We evaluate drone noise suppression performance using the following two metrics:

\textbf{\textit{Scale-Invariant Signal-to-Distortion Ratio (SI-SDR~\cite{le2019sdr})}}: SI-SDR measures how well the estimated source waveform, $\hat{s}$, reconstructs the clean ground-truth source $s$. We compute SI-SDR between the estimated source and the corresponding clean source-only recording for each configuration in DroneAudioSet~\cite{gupta2025droneaudioset}. To account for small temporal misalignments, we evaluate SI-SDR over shifts of up to $\pm$1000 samples ($\approx$62.5 ms at 16 kHz) and report the maximum value across shifts. 

\textbf{\textit{Classification Performance (F1-Score):}} While SI-SDR captures signal reconstruction fidelity, it does not directly measure perceptual quality or downstream task utility. Hence, we pass the recovered source estimate $\hat{s}$ through SSLAM~\cite{alex2025sslam}, a self-supervised audio classification model with state-of-the-art performance on AudioSet. SSLAM is used as a frozen classifier and applied identically across all evaluated methods. For each 5-second audio segment, SSLAM predicts the most likely class from the 527-class AudioSet ontology. Following the evaluation protocol of DroneAudioSet~\cite{gupta2025droneaudioset}, these predictions are mapped into four coarse categories: Human Vocal (HV), Human Non-Vocal (HNV), Non-Human (NH), and Not Detected/Silence (ND). We report macro F1-score, computed as the harmonic mean of precision and recall across the four categories.

\subsection{Implementation Details}
\label{sec:eval-model-details}
For \name's implementation, we utilize ResUNet from AudioSep, and utilize the same encoder/decoder blocks and FiLM conditioning. The main difference comes in the \texttt{after\_conv} projection layer, which converts the decoder's hidden representation into the actual prediction channels used to reconstruct audio. While the original ResUNet outputs three channels (corresponding to mask magnitude, as well as mask real and imaginary components, for obtaining the phase), \name's ResUNet outputs six channels, with three additional channels corresponding to residual magnitude, real and imaginary components. All the additional weights/biases for predicting the additional channels are randomly assigned using Xavier initialization~\cite{glorot2010understanding}. \name also has two learnable scalar components, $\alpha$ and $\beta$, both initialized to 1. Similar to AudioSep, we utilize L1-loss between the target and the estimated drone sounds, as our loss function. Furthermore, we decide the number of training epochs based on the validation loss. Rather than training \name from scratch, we fine-tune the publicly available AudioSep model pre-trained on the AudioSet dataset. Appendix section~\ref{app:computational-resources} elaborates on the computing and memory resources required to run our models.

\section{Results}
\label{sec:results}
\subsection{Overall Performance on DroneAudioset}
\label{sec:results-overall-performance}

\begin{table*}[t]
\centering
\scriptsize
\setlength{\tabcolsep}{4pt}
\renewcommand{\arraystretch}{1.05}
\resizebox{\textwidth}{!}{%
\begin{tabular}{l|c|ccc|cc}
\toprule
\textbf{SNR (dB)} 
& \textbf{Noisy} 
& \textbf{ZeroShot\cite{chen2022zero}} 
& \textbf{USS\cite{kong2023universal}} 
& \textbf{AudioSep\cite{liu2024separate}} 
& \textbf{AudioSep-FT} 
& \textbf{\name} \\
&&\multicolumn{3}{c|}{}&\multicolumn{2}{c}{}\\
&&\multicolumn{3}{c|}{\textbf{Without Finetuning}}&\multicolumn{2}{c}{\textbf{With Finetuning}}\\
\midrule

\multicolumn{7}{c}{\textbf{(a) SI-SDR (dB) $\uparrow$}} \\
\midrule

\rowcolor{gray!20}
\multicolumn{7}{c}{\textbf{Human Vocals (HV)}} \\
$-10$ to $0$ 
& \cellcolor[HTML]{FADBD8}$-11.87 \pm 5.21$
& \cellcolor[HTML]{FEF9E7}$-5.71 \pm 5.69$
& \cellcolor[HTML]{FEF9E7}$-6.41 \pm 6.34$
& \cellcolor[HTML]{E8F5E9}$-3.67 \pm 4.63$
& \cellcolor[HTML]{C8E6C9}{$\mathbf{-3.36 \pm 4.08}$}
& \cellcolor[HTML]{C8E6C9}\second{$-3.48 \pm 4.08$} \\

$-20$ to $-10$ 
& \cellcolor[HTML]{FADBD8}$-21.14 \pm 5.78$
& \cellcolor[HTML]{FDEBD0}$-17.85 \pm 7.74$
& \cellcolor[HTML]{FDEBD0}$-19.35 \pm 8.45$
& \cellcolor[HTML]{FEF9E7}$-13.88 \pm 8.95$
& \cellcolor[HTML]{C8E6C9}$\mathbf{-9.54 \pm 8.07}$
& \cellcolor[HTML]{C8E6C9}\second{$-9.71 \pm 8.11$} \\

$-30$ to $-20$ 
& \cellcolor[HTML]{FADBD8}$-26.87 \pm 3.14$
& \cellcolor[HTML]{FDEBD0}$-26.45 \pm 3.89$
& \cellcolor[HTML]{FDEBD0}$-26.66 \pm 3.95$
& \cellcolor[HTML]{FDEBD0}$-25.51 \pm 5.44$
& \cellcolor[HTML]{C8E6C9}$\mathbf{-21.64 \pm 8.36}$
& \cellcolor[HTML]{C8E6C9}\second{$-21.68 \pm 8.24$} \\

\midrule
\rowcolor{gray!20}
\multicolumn{7}{c}{\textbf{Human Non-Vocal (HNV)}} \\
$-10$ to $0$ 
& \cellcolor[HTML]{FADBD8}$-18.45 \pm 3.97$
& \cellcolor[HTML]{E8F5E9}$-9.45 \pm 5.34$
& \cellcolor[HTML]{FEF9E7}$-11.08 \pm 6.11$
& \cellcolor[HTML]{E8F5E9}$-8.52 \pm 5.16$
& \cellcolor[HTML]{C8E6C9}$\mathbf{-6.00 \pm 3.54}$
& \cellcolor[HTML]{C8E6C9}\second{$-6.14 \pm 3.53$} \\

$-20$ to $-10$ 
& \cellcolor[HTML]{FADBD8}$-24.97 \pm 3.50$
& \cellcolor[HTML]{FDEBD0}$-20.90 \pm 5.36$
& \cellcolor[HTML]{FDEBD0}$-21.65 \pm 6.73$
& \cellcolor[HTML]{FEF9E7}$-19.40 \pm 6.20$
& \cellcolor[HTML]{C8E6C9}$\mathbf{-12.25 \pm 7.18}$
& \cellcolor[HTML]{C8E6C9}\second{$-12.33 \pm 7.14$} \\

$-30$ to $-20$ 
& \cellcolor[HTML]{FADBD8}$-27.85 \pm 1.94$
& \cellcolor[HTML]{FDEBD0}$-26.13 \pm 2.46$
& \cellcolor[HTML]{FADBD8}$-27.18 \pm 3.35$
& \cellcolor[HTML]{FDEBD0}$-26.80 \pm 3.43$
& \cellcolor[HTML]{C8E6C9}$\mathbf{-23.13 \pm 6.77}$
& \cellcolor[HTML]{C8E6C9}\second{$-23.18 \pm 6.72$} \\

\midrule
\rowcolor{gray!20}
\multicolumn{7}{c}{\textbf{Non-Human Sounds (NH)}} \\
$-10$ to $0$ 
& \cellcolor[HTML]{FADBD8}$-7.38 \pm 4.32$
& \cellcolor[HTML]{E8F5E9}$0.39 \pm 6.12$
& \cellcolor[HTML]{FEF9E7}$-0.32 \pm 6.27$
& \cellcolor[HTML]{E8F5E9}$1.62 \pm 4.27$
& \cellcolor[HTML]{C8E6C9}$\mathbf{2.17 \pm 3.27}$
& \cellcolor[HTML]{C8E6C9}\second{$2.00 \pm 3.22$} \\

$-20$ to $-10$ 
& \cellcolor[HTML]{FADBD8}$-15.30 \pm 5.75$
& \cellcolor[HTML]{FDEBD0}$-10.75 \pm 8.57$
& \cellcolor[HTML]{FDEBD0}$-11.02 \pm 9.91$
& \cellcolor[HTML]{FEF9E7}$-6.54 \pm 7.46$
& \cellcolor[HTML]{C8E6C9}$\mathbf{-1.66 \pm 6.92}$
& \cellcolor[HTML]{C8E6C9}\second{$-1.84 \pm 6.88$} \\

$-30$ to $-20$ 
& \cellcolor[HTML]{FADBD8}$-25.10 \pm 4.41$
& \cellcolor[HTML]{FDEBD0}$-24.26 \pm 5.96$
& \cellcolor[HTML]{FDEBD0}$-24.76 \pm 6.11$
& \cellcolor[HTML]{FEF9E7}$-22.25 \pm 7.72$
& \cellcolor[HTML]{C8E6C9}\second{$-18.23 \pm 10.39$}
& \cellcolor[HTML]{C8E6C9}$\mathbf{-17.90 \pm 10.05}$ \\

\midrule
\midrule

\multicolumn{7}{c}{\textbf{(b) Classification F1-score $\uparrow$}} \\
\midrule

\rowcolor{gray!20}
\multicolumn{7}{c}{\textbf{Human Vocals (HV)}} \\
$-10$ to $0$ 
& \cellcolor[HTML]{FDEBD0}$0.41 \pm 0.36$
& \cellcolor[HTML]{FADBD8}$0.35 \pm 0.33$
& \cellcolor[HTML]{FEF9E7}$0.49 \pm 0.33$
& \cellcolor[HTML]{E8F5E9}$0.70 \pm 0.26$
& \cellcolor[HTML]{C8E6C9}\second{$0.85 \pm 0.15$}
& \cellcolor[HTML]{C8E6C9}$\mathbf{0.86 \pm 0.13}$ \\

$-20$ to $-10$ 
& \cellcolor[HTML]{FDEBD0}$0.25 \pm 0.32$
& \cellcolor[HTML]{FADBD8}$0.20 \pm 0.28$
& \cellcolor[HTML]{FEF9E7}$0.28 \pm 0.32$
& \cellcolor[HTML]{FEF9E7}$0.41 \pm 0.37$
& \cellcolor[HTML]{E8F5E9}\second{$0.66 \pm 0.31$}
& \cellcolor[HTML]{C8E6C9}$\mathbf{0.73 \pm 0.28}\sigft$ \\

$-30$ to $-20$ 
& \cellcolor[HTML]{FDEBD0}$0.05 \pm 0.17$
& \cellcolor[HTML]{FADBD8}$0.04 \pm 0.14$
& \cellcolor[HTML]{FDEBD0}$0.05 \pm 0.17$
& \cellcolor[HTML]{FDEBD0}$0.05 \pm 0.17$
& \cellcolor[HTML]{E8F5E9}\second{$0.13 \pm 0.28$}
& \cellcolor[HTML]{C8E6C9}$\mathbf{0.17 \pm 0.32}\sigft$ \\

\midrule
\rowcolor{gray!20}
\multicolumn{7}{c}{\textbf{Human Non-Vocal (HNV)}} \\
$-10$ to $0$
& \cellcolor[HTML]{FADBD8}$0.12 \pm 0.14$
& \cellcolor[HTML]{FDEBD0}$0.28 \pm 0.10$
& \cellcolor[HTML]{FDEBD0}$0.24 \pm 0.11$
& \cellcolor[HTML]{FEF9E7}$0.50 \pm 0.22$
& \cellcolor[HTML]{C8E6C9}$\mathbf{0.72 \pm 0.22}$
& \cellcolor[HTML]{C8E6C9}\second{$0.69 \pm 0.22$} \\

$-20$ to $-10$ 
& \cellcolor[HTML]{FADBD8}$0.06 \pm 0.12$
& \cellcolor[HTML]{FDEBD0}$0.17 \pm 0.18$
& \cellcolor[HTML]{FDEBD0}$0.22 \pm 0.20$
& \cellcolor[HTML]{C8E6C9}$\mathbf{0.50 \pm 0.30}$
& \cellcolor[HTML]{E8F5E9}$0.42 \pm 0.25$
& \cellcolor[HTML]{E8F5E9}\second{$0.44 \pm 0.25$} \\

$-30$ to $-20$ 
& \cellcolor[HTML]{FADBD8}$0.02 \pm 0.09$
& \cellcolor[HTML]{FDEBD0}$0.05 \pm 0.17$
& \cellcolor[HTML]{FEF9E7}\second{$0.14 \pm 0.25$}
& \cellcolor[HTML]{C8E6C9}$\mathbf{0.34 \pm 0.36}$
& \cellcolor[HTML]{FDEBD0}$0.11 \pm 0.19$
& \cellcolor[HTML]{FEF9E7}$0.13 \pm 0.21$ \\

\midrule
\rowcolor{gray!20}
\multicolumn{7}{c}{\textbf{Non-Human Sounds (NH) }} \\
$-10$ to $0$ 
& \cellcolor[HTML]{E8F5E9}\second{$0.58 \pm 0.24$}
& \cellcolor[HTML]{C8E6C9}$\mathbf{0.65 \pm 0.22}$
& \cellcolor[HTML]{FADBD8}$0.42 \pm 0.16$
& \cellcolor[HTML]{FDEBD0}$0.45 \pm 0.17$
& \cellcolor[HTML]{FDEBD0}$0.44 \pm 0.13$
& \cellcolor[HTML]{FADBD8}$0.40 \pm 0.18$ \\

$-20$ to $-10$ 
& \cellcolor[HTML]{E8F5E9}$0.39 \pm 0.27$
& \cellcolor[HTML]{FEF9E7}$0.37 \pm 0.29$
& \cellcolor[HTML]{FADBD8}$0.24 \pm 0.18$
& \cellcolor[HTML]{FDEBD0}$0.32 \pm 0.18$
& \cellcolor[HTML]{C8E6C9}$\mathbf{0.40 \pm 0.16}$
& \cellcolor[HTML]{FEF9E7}\second{$0.38 \pm 0.18$} \\

$-30$ to $-20$
& \cellcolor[HTML]{FDEBD0}$0.17 \pm 0.26$
& \cellcolor[HTML]{C8E6C9}$\mathbf{0.34 \pm 0.37}$
& \cellcolor[HTML]{FADBD8}$0.14 \pm 0.15$
& \cellcolor[HTML]{FDEBD0}$0.18 \pm 0.17$
& \cellcolor[HTML]{E8F5E9}\second{$0.24 \pm 0.14\sigft$}
& \cellcolor[HTML]{FEF9E7}$0.21 \pm 0.14$ \\

\bottomrule
\end{tabular}%
}
\caption{(a) SI-SDR (dB) and (b) downstream classification F1-scores across sound categories and SNR ranges. Values report mean $\pm$ standard deviation, with pastel heatmap (row-wise normalization). Best values are in bold and second-best values are underlined. $\dagger$ indicates statistically significant improvement in pairwise t-test comparison between AudioSep-FT and \name\ ($p < 0.05$), with the marker assigned to the better-performing model. }
\label{tab:sisdr_benchmark_heatmap}
\end{table*}

Table~\ref{tab:sisdr_benchmark_heatmap} reports SI-SDR and F1-score results across the three sound classes HV, HNV, and NH, and three input SNR ranges or bands, i.e.~$> -10$~dB, 
$-20$ to $-10$~dB, and $-30$ to $-20$~dB, to evaluate performance variation across acoustic conditions. The reported results are on the test data from DroneAudioset for the noisy mixture, noise suppressed source signals using the baseline models Zeroshot \cite{chen2022zero}, USS \cite{kong2023universal}, and AudioSep \cite{liu2024separate}, as well as the fine-tuned version of AudioSep (AudioSep-FT), and \name. 

\noindent{\textbf{AudioSep outperforms all other baselines.}}
Across both signal-level (SI-SDR) and downstream classification (F1-score) evaluations, AudioSep consistently emerges as the strongest baseline in comparison to the other two methods (ZeroShot and USS), in the zero-shot setting. It achieves the highest SI-SDR across most sound categories and SNR conditions, 
and similarly outperforms other baselines in classification performance for human vocal and human non-vocal sounds. 
However, all methods, including AudioSep, exhibit noticeable degradation in lower SNR conditions (below -10 dB), indicating that challenging noise regimes remain an open problem. Based on its consistent superiority across both reconstruction fidelity and downstream task performance, we adopt AudioSep as the base architecture for subsequent model development, while specifically targeting improvements in these lower SNR regimes.

\noindent{\textbf{Effect of Finetuning on DroneAudioset dataset.}}
Finetuning the base AudioSep model with drone-specific recordings yields substantial gains across both SI-SDR and downstream classification performance. The fine-tuned variant (AudioSep-FT) consistently improves over the zero-shot baselines across all sound categories, with particularly pronounced gains in low and mid SNR regimes (below -10 dB), where the drone noise 
is most severe. This trend is reflected not only in improved signal reconstruction fidelity (through SI-SDR), but also in higher F1-scores for human vocal and human non-vocal categories, indicating better recovery of semantically meaningful content. These results highlight the importance of domain-specific finetuning in handling drone noise characteristics, which differ significantly from the training distributions of generic audio separation models.

\noindent{\textbf{Effect of Mask Parameterization and Residual Correction.}}
Building on the fine-tuned baseline, \name refines mask estimation through a relaxed sigmoid formulation together with explicit complex residual correction (Section~\ref{sec:methods-improved-mask-estimation}). While SI-SDR improvements over AudioSep-FT remain modest and are generally comparable across most sound categories and SNR regimes, \name achieves consistent gains in downstream classification performance for Human Vocals (HV), with statistically significant improvements in the challenging low-SNR conditions (10.6\% relative improvement over AudioSep-FT in $-10$ to $-20$ dB SNR range and 30.8\% in $-20$ to $-30$ dB SNR range). These gains suggest that the proposed mask parameterization better preserves semantically discriminative components relevant for downstream recognition, even when waveform-level reconstruction improvements are limited. Performance for Human Non-Vocal (HNV) and Non-Human (NH) sounds remains broadly comparable to AudioSep-FT, indicating that the primary benefit of the proposed formulation is in recovering structured and harmonically rich signals such as speech and vocal distress sounds under severe drone interference. Complementing the above F1-score results, Appendix Figure~\ref{fig:confusionmatrices} further illustrates confusion matrices for AudioSep, AudioSep-FT and \name.

\subsection{Ablation Study}
\label{sec:results-ablation}

\begin{table*}[t]
\centering
\scriptsize
\setlength{\tabcolsep}{4pt}
\renewcommand{\arraystretch}{1.05}
\resizebox{1.0\textwidth}{!}{%
\begin{tabular}{lccccccc}
\toprule
\textbf{Model} & $\alpha$ & $\beta$ & $|R|$ & $\angle R$ & \textbf{$\geq -10$ dB} & \textbf{$-20$ to $-10$ dB} & \textbf{$< -20$ dB} \\
\midrule


AudioSep-FT
& \xmark
& \xmark
& \xmark
& \xmark
& \cellcolor[HTML]{E8F5E9}$0.85 \pm 0.15$ 
& \cellcolor[HTML]{E8F5E9}$0.66 \pm 0.31$ 
& \cellcolor[HTML]{E8F5E9}$0.13 \pm 0.28$ 
\\

AudioSep-FT w/ mask scaling parameter $\alpha$
& $\alpha=1.0294$
& \xmark
& \xmark
& \xmark
& \cellcolor[HTML]{C8E6C9}$0.86 \pm 0.14$ 
& \cellcolor[HTML]{C8E6C9}$\mathbf{0.73 \pm 0.28}$ 
& \cellcolor[HTML]{C8E6C9}$0.16 \pm 0.32$ 
\\

AudioSep-FT w/ $\alpha$, magnitude residual 
& $\alpha=1.0342$
& $\beta=0.8264$
& \cmark
& \xmark
& \cellcolor[HTML]{FEF9E7}$0.75 \pm 0.20$ 
& \cellcolor[HTML]{FEF9E7}$0.60 \pm 0.32$ 
& \cellcolor[HTML]{E8F5E9}$0.13 \pm 0.28$ 
\\

\name w/ $\alpha$, complex residual
& $\alpha=1.0196$
& $\beta=0.9743$
& \cmark
& \cmark
& \cellcolor[HTML]{C8E6C9}$\mathbf{0.86 \pm 0.13}$ 
& \cellcolor[HTML]{C8E6C9}$\mathbf{0.73 \pm 0.28}$ 
& \cellcolor[HTML]{C8E6C9}$\mathbf{0.17 \pm 0.32}$ 
\\

\bottomrule
\end{tabular}%
}
\caption{Ablation study using Classification F1-score metric for human-vocals (HV) (pastel heatmap with column-wise normalization).}
\label{tab:ablation}
\end{table*}









\begin{table}[t]
\centering
\scriptsize
\setlength{\tabcolsep}{5pt}
\renewcommand{\arraystretch}{1.1}
\begin{tabular}{lccc}
\toprule
\textbf{} 
& \textbf{AudioSep} 
& \textbf{AudioSep-FT} 
& \textbf{\name} \\
\midrule


\textbf{Human Vocals} 
& \cellcolor[HTML]{FDE0DC}$0.50 \pm 0.29$
& \cellcolor[HTML]{EAF4E6}$0.63 \pm 0.27$
& \cellcolor[HTML]{C8E6C9}$\mathbf{0.69 \pm 0.25}\sigft$ \\

\midrule

\textbf{Human Non-Vocals} 
& \cellcolor[HTML]{EAF4E6}$\mathbf{0.26 \pm 0.19}$
& \cellcolor[HTML]{F3F6E5}$0.25 \pm 0.20$
& \cellcolor[HTML]{F3F6E5}$0.25 \pm 0.19$ \\

\midrule

\textbf{Non-Human Sounds} 
& \cellcolor[HTML]{FDE0DC}$0.31 \pm 0.21$
& \cellcolor[HTML]{EAF4E6}$0.46 \pm 0.20$
& \cellcolor[HTML]{C8E6C9}$\mathbf{0.48 \pm 0.22}$ \\

\bottomrule
\end{tabular}
\caption{Out-of-domain classification performance (F1-score) on DREGON at $-20$ to $-10$~dB SNR. Values are reported as mean $\pm$ standard deviation (row-wise normalized heatmap). $\dagger$ denotes statistically significant improvement ($p<0.05$) in pairwise t-tests between AudioSep-FT and \name, assigned to the better-performing model.}

\label{tab:dregon_ood_final}
\end{table}

We perform an ablation study to analyze the contribution of the learnable mask-scaling parameter $\alpha$ and the residual correction term $\beta |R|e^{j\angle R}$ in \name. Results are reported using downstream classification F1-score for human vocal sounds across different SNR regimes in Table~\ref{tab:ablation}.

\noindent{\textbf{Effect of learnable mask scaling ($\alpha$).}}
Comparing AudioSep-FT with the variant that introduces only the learnable scaling parameter $\alpha$ (without residual correction) shows consistent improvements across all SNR regimes, with significant improvement in $-20$ to $-10$ dB and $-30$ to $-20$ dB. This suggests that relaxing the bounded mask through $\alpha$ improves drone-noise estimation when the interference strongly dominates the mixture. Notably, although $\alpha$ is initialized to $1$, the model consistently learns values greater than $1$, indicating that the network naturally benefits from scaling the mask beyond the unit-circle constraint imposed by the sigmoid formulation. This supports our hypothesis that standard sigmoid-constrained masks are overly restrictive for noise-dominant mixtures. Appendix Figure~\ref{fig:predicted-masks} illustrates the masks learnt by the AudioSep-FT and \name models, to further demonstrate the influence of parameter, $\alpha$.  

\noindent{\textbf{Effect of residual correction.}}
Adding only the residual magnitude term without phase prediction leads to a noticeable performance drop across all SNR regimes, indicating that magnitude-only correction is insufficient. In contrast, incorporating the full complex residual, including both $|R|$ and $\angle R$, restores and slightly improves performance, particularly in the lowest SNR regime. Interestingly, the learned residual scaling parameter remains below its initialization value of $1$, suggesting that the residual branch acts primarily as a complementary corrective refinement rather than a dominant reconstruction pathway. These findings indicate that the residual phase component provides a small but useful corrective signal beyond the scaled mask estimate.

\subsection{Out-of-domain generalization on DREGON.}
\label{sec:results-ood}
We evaluate out-of-domain generalization on DREGON in the challenging -20 to -10 dB SNR regime, where 
the target distribution differs from training data. As shown in Table~\ref{tab:dregon_ood_final}, \name maintains competitive or improved performance compared to both AudioSep and AudioSep-FT. Gains are most pronounced for Human Vocals (0.69 vs. 0.63/0.50) and Non-Human Sounds (0.48 vs. 0.46/0.31), indicating improved recovery of semantically meaningful events under severe noise mismatch. In contrast, performance on Human Non-Vocals remains comparable across methods, suggesting that these signals, being less structured and often transient, are inherently harder to recover in low-SNR, out-of-domain settings. Overall, these results highlight the robustness of \name\ to distribution shift, particularly for acoustically salient and structured sources.

\begin{figure}
    \centering
    \includegraphics[width=0.7\linewidth]{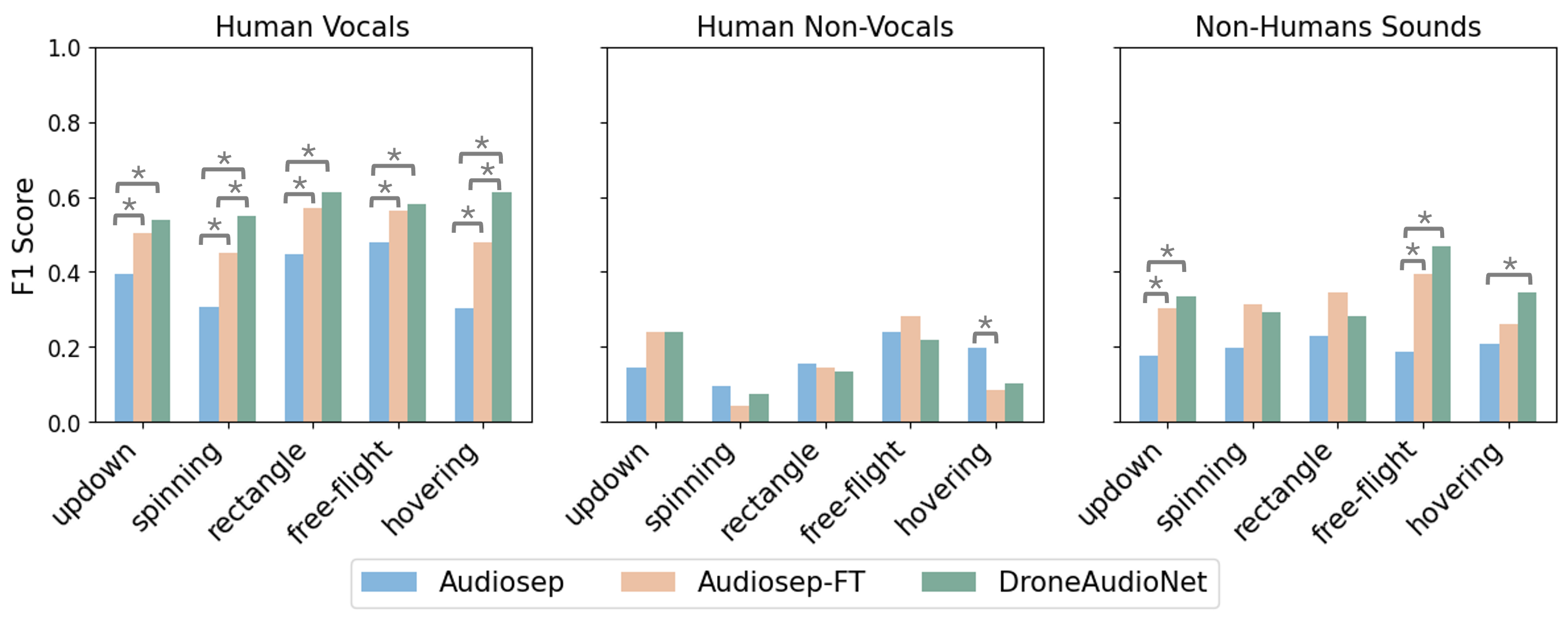}
    \caption{Dregon OOD: F1 Score by Drone Mode and Model (per Class). The * refers to statistically significant difference ($p<0.05$) in a pairwise t-test between two conditions.}
    \label{fig:dregon_barplot}
\end{figure}

\textit{Generalisation across drone flight modes.} 
Figure 3 further breaks down performance across drone flight modes, showing consistent advantages of \name\ across diverse acoustic conditions. Improvements are especially evident in complex modes such as \emph{free-flight}, where noise characteristics are highly non-stationary and differ substantially from training conditions. The strong performance observed in \emph{hovering} can be attributed to its alignment with the data distribution used in DroneAudioSet, where hovering is the primary recording configuration. Notably, \name\ achieves higher F1 scores across nearly all modes for Human Vocals and Non-Human Sounds, with statistically significant gains in several cases. This suggests that explicitly modeling drone noise with an unbounded complex mask and residual correction enables better generalization to unseen environments, particularly in challenging settings like free-flight.

\vspace{-0.3cm}
\section{Discussion}
\label{sec:discussion}
\vspace{-0.3cm}
\noindent{\textbf{Practical Implications for Drone-based Search and Rescue.}}
Our results suggest that explicit drone-noise modeling is valuable for search-and-rescue scenarios where rotor noise dominates ambient sounds. The strongest gains are observed in the mid-SNR regime, 
particularly for human vocal sounds relevant to distress detection. 
Additionally, strong out-of-domain performance on DREGON indicates robustness to unseen drone hardware and flight modes. However, practical deployment still requires low-latency onboard inference, improved robustness to environmental noise, and validation with real-world rescue personnel and operating conditions.

\noindent{\textbf{Broader Impacts and Safeguards.}}
This work has potential positive impact for drone-assisted search-and-rescue, disaster response, wildlife monitoring, and remote environmental sensing by improving aerial acoustic perception in visually challenging conditions. However, drone-mounted microphones also raise concerns regarding privacy, surveillance, and misuse, while incorrect predictions in extreme low-SNR conditions may lead to false alarms or missed detections. Accordingly, \name should be viewed as an assistive perception module requiring human oversight rather than a fully autonomous decision-making system.

We release \name as a research benchmark for drone-noise suppression rather than a surveillance pipeline, and the system does not perform identity recognition, speaker profiling, or long-term audio monitoring. We encourage future deployments to comply with relevant privacy, aviation, and ethical regulations for aerial sensing technologies.

\noindent{\textbf{Limitations and Future Work}}
DroneAudioSet primarily contains hovering recordings, whereas real-world deployments involve dynamic flight trajectories, wind, and environmental reverberation; broader real-world evaluation is therefore needed. In addition, our framework operates on single-channel audio and does not exploit spatial information from multi-microphone arrays. Finally, performance under extremely adverse SNR conditions and for transient acoustic events remains challenging, motivating future work on 
end-to-end enhancement-and-detection frameworks.

\vspace{-0.3cm}
\section{Conclusion}
\vspace{-0.3cm}
In this work, we presented \name, a drone noise suppression method designed for drone-based search-and-rescue scenarios. Building upon the state-of-the-art source separation framework AudioSep, we adapted the model to the acoustic characteristics of drone noise through a learnable mask-scaling mechanism and an additive residual branch for improved drone noise estimation and source recovery. Our results demonstrate that fine-tuning on drone-specific recordings substantially improves both source reconstruction and downstream classification performance, while the proposed architectural modifications further improve downstream classification performance for human vocal sounds in challenging low-SNR and out-of-domain settings.

\bibliographystyle{plain}
\bibliography{paper}



\appendix
\section{Extended Analysis and Results}
\label{app:extended-results}
\subsection{Webpage, Codebase and Models}
\label{app:webpage}
\begin{itemize}
    \item Webpage (anonymized): for listening to example outputs
\url{https://droneaudionet.github.io/DroneAudioNet-Homepage/}
\item Codebase (anonymized): \url{https://github.com/DroneAudioNet/DroneAudioNet-Homepage}
\end{itemize}

\begin{table}[]
\renewcommand{\arraystretch}{1}
\resizebox{\textwidth}{!}{
\begin{tabular}{|l|l|l|l|l|l|}
\hline

& Exec Time (min)
& Max CPU Usage (\%)
& Memory Usage (MB)
& Max GPU Usage (\%)
& GPU Memory Usage (MB)       \\ \hline

Training
& \multicolumn{1}{c|}{90.5}
& \multicolumn{1}{c|}{0.0}
& \multicolumn{1}{c|}{49440.3}
& \multicolumn{1}{c|}{100 (x2)}
& \multicolumn{1}{c|}{12241.6} \\ \hline
Inference
& \multicolumn{1}{c|}{23.5}
& \multicolumn{1}{c|}{2.9}
& \multicolumn{1}{c|}{3007.4}
& \multicolumn{1}{c|}{99}
& \multicolumn{1}{c|}{1602.4} \\ \hline

\end{tabular}}
\caption{Computational resources required for training and inference to process the training and test dataset from DroneAudioset, respectively. For training, we report execution time per epoch.  }
\label{tab:computational-resources}
\end{table}

\subsection{Computational Resources for Evaluation}
\label{app:computational-resources}
Our \name model has a total of 238M parameters, with 212M non-trainable parameters and 26.4M trainable parameters. This amounts to a total memory requirement of 954.4 MB. Majority of the non-trainable parameters belong to the CLAP encoder~\cite{wu2023large} which we use to obtain the embedding corresponding to the natural language query. The entire ResUNet model belongs under the trainable parameters. 

All experiments were conducted on a Linux system (5.15.0-119-generic) with an x86\_64 architecture using Python 3.9.21. The hardware configuration consisted of a 36-thread CPU (18 physical cores @ 2.95 GHz), 134.7 GB of system RAM, and an NVIDIA GeForce RTX 3090 GPU with 24GB of VRAM (utilizing 58MB during idle measurements at 28°C). The computational resources required at each stage of the benchmarking pipeline is provided in Table \ref{tab:computational-resources}.

\begin{figure}
    \centering
    \includegraphics[width=0.7\linewidth]{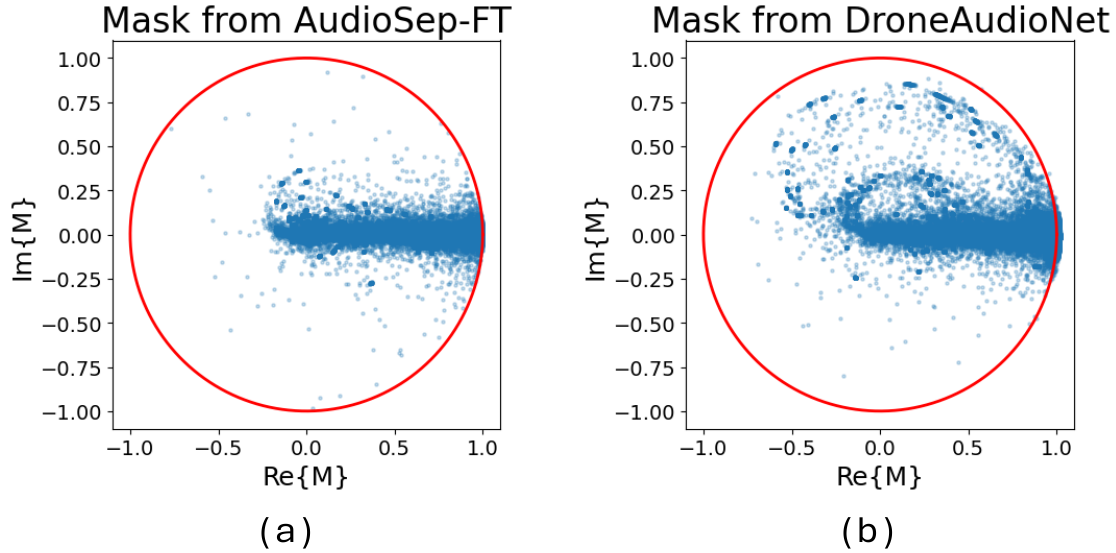}
    \caption{Masks predicted from the (a) AudioSep-FT and (b) \name models, with the unit circle shown in red. While 23.6\% of mask values lie outside the unit circle, none of the values from AudioSep-FT exceed unity, demonstrating the influence of the $\alpha$ parameter in \name. }
    \label{fig:predicted-masks}
\end{figure}

\subsection{Classification Confusion Matrices}
\label{app:confusionmatrices}

\begin{figure}
    \centering
    \includegraphics[width=\linewidth]{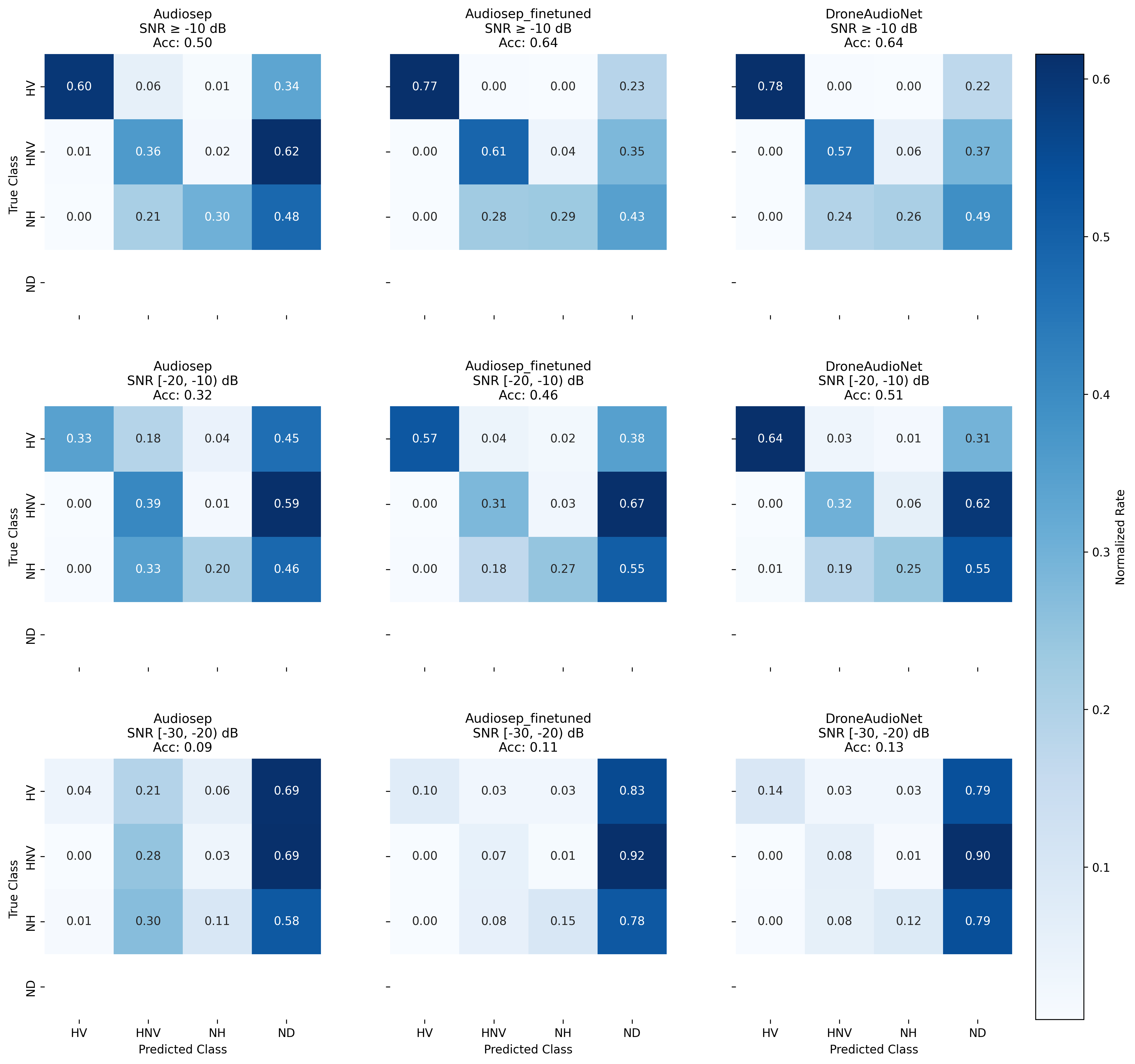}
    \caption{Confusion Matrices of the three models across the three input SNR bands.}
    \label{fig:confusionmatrices}
\end{figure}

Figure~\ref{fig:confusionmatrices} shows normalized confusion matrices across the three SNR bands for AudioSep, AudioSep-FT, and \name. As SNR decreases, all methods increasingly confuse source sounds with the ND (Not Detected) category, reflecting the difficulty of recovering weak signals under dominant drone noise. Fine-tuning substantially improves recognition of Human Vocal (HV) sounds compared to the zero-shot AudioSep baseline, particularly in the $-20$ to $-10$ dB regime. \name further improves HV recovery in low-SNR conditions while maintaining comparable performance for Human Non-Vocal (HNV) and Non-Human (NH) sounds. These trends are consistent with the quantitative F1-score improvements reported in Table~\ref{tab:sisdr_benchmark_heatmap}.

\subsection{Predicted Mask Visualization}
\label{app:mask}
In Section~\ref{sec:methods-improved-mask-estimation}, we explain the need for mask magnitude greater than unity for drone noise estimation. Hence, in Figure~\ref{fig:predicted-masks}, we depict the masks predicted by the two models -- AudioSep-FT and \name. As shown, while mask estimation of AudioSep-FT makes it impossible to learn masks with magnitude over 1 (given the \texttt{sigmoid}) constraint, we observe that \name's predicted mask has 23.6\% of its values beyond 1, demonstrating the effect of the $\alpha$ ($=1.0196$). Additionally, while most predictions from AudioSep-FT lie close to the real axis, \name produces substantially more non-zero imaginary components, indicating that the model learns meaningful phase corrections.



\end{document}